\documentstyle[sprocl]{article}
\def\simlt{\stackrel{<}{{}_\sim}}
\def\simgt{\stackrel{>}{{}_\sim}}
\def\MK{M}
\def\MP{M_{Pl}}
\def\MS{M_{string}}
\def\9{\phantom 0}
\renewcommand\linebreak{\unskip\break} 
\def\lsim{\lower 3pt\hbox{$\buildrel<\over\sim$}}

\def\to{\rightarrow}

\def\etal{{\it et al.}}
\newcommand{\ZP}[3]{Zeit. Phys. {\bf #1}, #2 (19#3)}                            
\newcommand{\PR}[3]{Phys. Rev. {\bf #1}, #2 (19#3)}                             
\newcommand{\PL}[3]{Phys. Lett. {\bf #1}, #2 (19#3)}                            
\newcommand{\NP}[3]{Nucl. Phys. {\bf #1}, #2 (19#3)}                            
                      
\newcommand{\Rep}[3]{Phys. Rep. {\bf #1}, #2 (19#3)}

\newcommand{\con}[3]{ {\bf #1}, #2 (19#3)}                                      
                                 
\begin{document}
\hfill{UPR-0761-T, hep-ph/9707451}
\vspace{.5cm}
\title{$Z'$ Physics and Supersymmetry\footnote{To appear
in {\it Perspectives in Supersymmetry}, World Scientific, ed. G. L. Kane.}}
\author{M. Cveti\v c and  P. Langacker}
\address{Department of Physics and Astronomy, University of  
Pennsylvania, \\ Philadelphia,
PA 19104-6396}
\maketitle
\centerline{\small (\today)}
\vspace{.5cm}
\abstracts{
We review the status of heavy neutral gauge  bosons,  $Z'$, with   
emphasis on  constraints that arise  in supersymmetric models,
especially those motivated from superstring compactifications. We  
first summarize the current phenomenological constraints
and the prospects for $Z'$ detection and diagnostics at the
LHC and NLC. After
elaborating on the  status and (lack of) predictive power for  
general   models with an additional $Z'$, we  concentrate on   
motivations  and successes for   $Z'$ physics in  supersymmetric  
theories in general and in a class of superstring models in  
particular. We review  phenomenologically viable scenarios with  
the $Z'$ mass  in the
electroweak  or in the  intermediate scale region.}

\setcounter{footnote}{0}
\newpage
\title{$Z'$ Physics and Supersymmetry}
\author{M. Cveti\v c and  P. Langacker}
\address{Department of Physics and Astronomy, University of  
Pennsylvania, \\ Philadelphia,
PA 19104-6396}
\maketitle\abstracts{
We review the status of heavy neutral gauge  bosons,  $Z'$, with   
emphasis on  constraints that arise  in supersymmetric models,
especially those motivated from superstring compactifications. We  
first summarize the current phenomenological constraints
and the prospects for $Z'$ detection and diagnostics at the
LHC and NLC. After
elaborating on the  status and (lack of) predictive power for  
general   models with an additional $Z'$, we  concentrate on   
motivations  and successes for   $Z'$ physics in  supersymmetric  
theories in general and in a class of superstring models in  
particular. We review  phenomenologically viable scenarios with  
the $Z'$ mass  in the
electroweak  or in the  intermediate scale region.}

\section{Introduction}
The existence of heavy neutral ($Z'$)  vector
gauge bosons are a feature of many extensions of the standard model  
(SM). In particular,   one (or more) additional  $U(1)'$  gauge  
factors  provide  one of the simplest extensions of the  
SM.
Additional $Z'$ gauge bosons  appear  in grand unified theories  
(GUT's), superstring  compactifications, and other classes of models.

However, for those models 
which do not incorporate constraints
due to  supersymmetry,
supergravity, or superstring theory,  the masses  
of additional gauge bosons are generally
free parameters which can range  from the  
electroweak scale ($\cal O$(1) TeV)  to the Planck scale $\MP$.  
In addition, models with extended gauge symmetry
generically  contain  exotic particles,
{e.g.}, new heavy quarks or leptons which are non-chiral under
$SU(2)_L$, or new  SM singlets,
with masses that are tied to those of the new $Z'$s,  but
are otherwise unconstrained.
Thus,  such  models  lack  predictive power for
$Z'$  physics, and
much of the phenomenological work  in this context is of the 
``searching under the lamp-post''
variety.  In particular, there is no strong motivation to think that
an extra $Z'$ would actually be light enough to be observed at future  
colliders.
Even within ordinary GUT's, there is no robust prediction for the
mass of a $Z'$. (There  {\em are}, however,
concrete predictions for the relative couplings of the ordinary
and exotic particles to the $Z'$ in each ordinary or  supersymmetric GUT.)

On the other hand   $N=1$ supersymmetric models typically provide  
more constraints.  First, the scalar potential is  determined by the  
superpotential, K\" ahler potential, $D$-terms, 
and soft supersymmetry breaking  
terms, and is generally more restrictive.  In particular, the $U(1)'$ 
$D$-term,  which
is typically of the order of $M^2_{Z'}$, breaks supersymmetry,
so the $U(1)'$ breaking scale should
not be much larger than the  electroweak scale,
i.e.,  no more
than a TeV or so~\cite{dterms,lykken}.
The  exception is  when
the  $U(1)'$ breaking  takes place along a $D$-flat direction~\footnote{It
was claimed in~\cite{lykken} that the breaking {\em must} be along a $D$-flat
direction. In fact, that is only necessary when the $U(1)'$ breaking 
scale is much larger than the supersymmetry breaking (i.e., TeV) scale.}.  In  
this case the breaking  can take place at an intermediate
scale~\footnote{The actual minimum is typically shifted slightly away from the
$D$-flat direction by soft supersymmetry breaking terms, leading to finite
shifts in the sparticle and Higgs masses of order of the soft breaking,
even for a large $U(1)'$  breaking scale.}.  Secondly, superstring models
often imply constraints on the superpotential, such as the absence
of mass terms and the existence of large (order one) Yukawa couplings,
which can determine the mechanism and scale of $U(1)'$  breaking.

There are two promising classes of theoretical structures in which the 
minimal supersymmetric standard model (MSSM) and its extensions are likely to 
be embedded.
One  are superstring models which compactify directly to
a gauge group consisting of the SM  and possibly additional
$U(1)'$ factors~\cite{SY,CL,CDEEL,CCEEL}.  Some of the  superstring
models are based on $E_6\times E_8$
Calabi-Yau compactifications of the heterotic string,
in which $E_6$ is already broken at the string scale, e.g.,
to the  SM 
gauge group and an additional $U(1)'$ factor, via the
Wilson line (Hosotani) mechanism.
Such $Z'$ models, primarily employing the quantum numbers
associated with a particular  $E_6$ breaking pattern, 
 were  addressed in~\cite{GRH,keithma,jhtr}.
 The  $Z'$ phenomenology  of  superstring models with  a 
 true GUT symmetry at the string scale has not been explored.  
Most of the recent work on
supersymmetric $Z'$s has been from the point of  view of the 
 first type, 
non-GUT models, and it will 
 be emphasized in this contribution.

In the models studied in~\cite{SY,CL,CDEEL,CCEEL,keithma}
the $U(1)'$  breaking is radiative. This is analogous to
radiative electroweak breaking, in which the (running)
Higgs mass-squared terms are positive
at the Planck scale, but one of them 
is driven negative at a low or intermediate scale by the large Yukawa coupling
associated with the top quark. Similarly,
in radiative $U(1)'$  breaking one or more SM scalars that carry  $U(1)'$
charges have positive mass-squared terms at the Planck scale. However, if these
scalars have
large Yukawa couplings to exotic multiplets or to Higgs doublets
their mass-squared terms can be
driven negative at  lower scales so that the scalar develops a
vacuum expectation value and breaks the $U(1)'$ symmetry.
Typically, the initial (Planck scale)
values of the Higgs and SM scalar mass-squared terms
are comparable and given by the scale of soft supersymmetry breaking.
 In a class of models in which the  magnitudes of the Yukawa couplings in the  
superpotential are motivated to be of $\cal O$(1), the radiative breaking  
can occur, and it  depends on the exotic particle content 
and  on the  boundary  
conditions for the soft supersymmetry
breaking terms at the large scale. The symmetry breaking can
either take place at the  electroweak~\cite{SY,CL,CDEEL,keithma}
or (when the minimum occurs along
a $D$-flat direction) at an intermediate  
scale~\cite{iscales,SY,CL,CCEEL}. 
In the former case, the $Z'$ mass
is comparable to the ordinary $Z$ and to the scale of supersymmetry breaking,
 and is almost certainly less than
a TeV~\cite{CL,CDEEL}.

A class of  
superstring compactifications, based on free fermionic
constructions~\cite{ABK,NAHE,faraggi90a,CHL,dienes}~\footnote{Certain
models based on orbifold constructions  with Wilson
lines\cite{INQ}  also possess the gauge structure and the particle content
of the MSSM.},  
contains all of these ingredients, including 
the general particle content and  gauge group 
of the MSSM,  and in  
general additional non-anomalous 
$U(1)'$ gauge symmetry factors and vector
pairs of exotic  chiral supermultiplets~\footnote{Related classes  of models
based on higher level Ka\v c-Moody algebra constructions
yield~\cite{CHL,AFIU,Kakushadze}
 GUT gauge symmetry with adjoint representations. These models have not yet been
 explored for $Z'$ physics.}.  
In this class of  models there are no bilinear (mass) terms
in the superpotential, and 
the  non-vanishing trilinear (Yukawa) terms are  of  
order one. These  conditions 
 usually suffice to
 require and allow  radiative $U(1)'$ breaking, respectively~\cite{CL}.  
Thus, supersymmetric models (with 
constraints on couplings from  a class of superstring models)  {\it  
 have predictive power}  for the  
masses of $Z'$ and the accompanying exotic particles. In that  sense they are
superior to  general models with  extended gauge  
structures. 
For these reasons, we would  like to advocate that within supersymmetry (and  
superstring theory constraints),  perhaps  the best motivated physics
for future experiments, next to the Higgs and sparticle
searches,  are searches for $Z'$ and the accompanying exotic  
particles.
On the other hand, at present we have little theoretical
control over the type of dynamically preferred
superstring compactification, or of the soft supersymmetry breaking 
parameters, i.e., the origin of supersymmetry breaking in
superstring models; it is therefore 
hard to make general predictions for the specific
$U(1)'$ charges and other details of the models.

In  supersymmetric models  additional $U(1)'$s would have  
important theoretical implications.
For example, an extra $U(1)'$ breaking at the electroweak scale
in a supersymmetric extension of the SM could
solve the $\mu$ problem~\cite{MUPROB}
by forbidding an elementary
$\mu$-parameter,
but inducing an effective $\mu$  of the order of the electroweak scale
by the $U(1)'$ breaking~\cite{SY,CL,CDEEL}.
There are also implications
for baryogenesis; an extra $U(1)'$ might be useful for electroweak
baryogenesis, with changes in the scalar sector of such models modifying the 
nature of the electroweak transition~\cite{CDEEL,esp}, or
 with cosmic strings providing the needed
``out of equilibrium'' ingredient~\cite{COSTRINGS}. However, alternative
models of baryogenesis, in which a lepton asymmetry is first created
by the out of equilibrium decay of a heavy Majorana neutrino, and
then converted to a baryon asymmetry by electroweak effects~\cite{leptonasym},
is forbidden by a light $Z'$ unless the  heavy neutrino carries no 
$U(1)'$ charge~\cite{buch}. Other  implications include the Higgs,
scalar partner, chargino,  and
neutralino masses and couplings, and thus
the properties of the LSP~\cite{decarlosespin}.
Intermediate scale breaking models may,
through higher-dimensional operators in the superpotential, yield an
attractive mechanism for generating hierarchies of small masses
for quarks and charged leptons, and offer possibilities for generating
naturally small Dirac neutrino masses without invoking a seesaw~\cite{CCEEL}.
Small Majorana neutrino masses (comparable to the  small 
Dirac neutrino masses), which allow oscillations of ordinary
into sterile neutrinos, 
or  large Majorana masses, generating  seesaws, are also possible. 
Non-renormalizable terms may also provide a mechanism for obtaining 
the  $\mu$-parameter.

This review is organized as follows.
In Section 2 we review phenomenological constraints on new $Z'$ bosons
from current precision measurements and collider experiments, the 
discovery potential of  future  
collider experiments, and  their diagnostic power for $Z'$  
identification.
In Section 3, we  review the  features of $Z'$  models with or without
supersymmetry  and with or without GUT embedding.  
In Section 4 we review the properties of  non-GUT supersymmetric models  
with additional $U(1)'$ factors,  address
 additional constraints and inputs arising  
in a class of superstring models, and discuss  both the electroweak  
and the intermediate scale scenarios for $U(1)'$ symmetry breaking.
Brief conclusions are given in Section 5.

\section{$Z'$ Physics}
In this Section we discuss the phenomenological  
constraints on new gauge bosons at current and future colliders.
For a more detailed analysis of $Z'$ physics at colliders,
 see~\cite{CG};
for constraints from
precision experiments, see~\cite{jepl}.

\subsection{Overview of $Z'$ Models}

Here,  we briefly describe some
examples of extended gauge theories.  While not
comprehensive, the properties are representative of models with extra gauge 
bosons. For a more detailed discussion, see~\cite{dl,jhtr}.

The most commonly studied $Z'$ couplings are GUT, left-right, and
string-motivated. In each case, the $Z'$ couples to
$g' Q$, with $g'$  the $U(1)'$ gauge coupling and $Q$
 the charge. 
In the GUT-motivated cases 
$g=\sqrt{5/3}\sin\theta_W G\lambda_g^{1/2}$, with 
$G\equiv\sqrt{g_L^2+g_Y^2}=g_L/\cos\theta_W$, where
 $g_L$, $g_Y$ are the
gauge couplings of $SU(2)_{L}$ and $U(1)_{Y}$, and
$\lambda_g$
depends on the symmetry breaking pattern~\cite{robinett}.
If the GUT group breaks
directly to $SU(3)_C\times SU(2)_L\times U(1)_Y\times U(1)'$,
 then $\lambda_g=1$.
Standard examples include: (i)
$Z_\chi$, which occurs in $SO_{10}\rightarrow SU(5)\times U(1)_{\chi}$;
(ii) $Z_\psi$, from $E_6\rightarrow SO_{10}\times U(1)_{\psi}$;
(iii) $Z_\eta=\sqrt{3/8}Z_\chi-\sqrt{5/8}Z_\psi$, which occurs in
Calabi-Yau  compactifications  of the heterotic string if
$E_6$ breaks directly to a rank 5
group~\cite{etamodel} via  the Wilson line (Hosotani) mechanism; (iv) 
the general $E_6$ boson $Z(\theta_{E_6})=\cos\theta_{E_6}
Z_\chi+\sin\theta_{E_6} Z_\psi$,
where $0\le\theta_{E_6}<\pi$ is a mixing angle.
The $Z_\chi$, $Z_\psi$, and $Z_\eta$ are special cases with
$\theta_{E_6}=0,{\pi\over2}$ and $Z_\eta=-Z$
($\theta_{E_6}=\pi-\arctan\sqrt{5/3}$), respectively; 
(v)
$Z_{LR}$ occurs in left-right symmetric (LR) models~\cite{lrmodels},
which contain a right-handed charged boson as well as an additional neutral
boson.
The ratio
$\kappa=g_R/g_L$ of the gauge couplings $g_{L,R}$ for $SU(2)_{L,R}$,
respectively, parametrizes the whole class of models.
$\kappa=1$
corresponds to manifest  or pseudo-manifest
left-right symmetry.
In this case $\lambda_g = 1$ by construction and $\kappa > 0.55$
for consistency~\cite{amaldi,kayser}. The $U(1)'$ 
charges for specific models are given in
Table~\ref{zprime}

\begin{table}[ht]
\caption[]{Couplings of the $Z_{\chi}$, $Z_{\psi}$, and
 $Z_{\eta}$ to a 27-plet of $E_6$. The $SO(10)$ and $SU(5)$
 representations are also indicated. The couplings are shown for
 the left-handed ($L$) particles and antiparticles. The couplings of
 the right-handed particles are minus those of the corresponding
 $L$ antiparticles. The $D$ is an exotic $SU(2)_L$-singlet quark
 with charge $-1/3$. $(E^{0}, E^{-})_{L,R}$ is an exotic lepton
 doublet with vector $SU(2)_L$ couplings. $N$ and $S$ are new Weyl
 neutrinos which may have large Majorana masses.
The $Z_{LR}$ couples to the charge $\sqrt{3/5}
\left[ \delta T_{3R}-1/(2\delta)
T_{B-L} \right] $, where $\delta=\sqrt{\kappa^2\cot^2\theta_W-1}$ and
$\kappa=g_R/g_L$.} \label{zprime} \centering
\begin{tabular}{|ccccc|} \hline
  $ SO(10) $  & $ SU(5) $ & $ 2 \sqrt{10} Q_{\chi} $ &
  $ \sqrt{24} Q_{\psi} $ & $ 2 \sqrt{15} Q_{\eta} $ \\
\hline
16  & 10 $(u,d,\bar{u},e^{+})_{L}$  & $-1$ & $1$ & $-2$ \\
    & $ 5^{*} (\bar{d},\nu, e^{-})_{L}$  & $3$ & $1$ & $1$ \\
    & $ 1 \bar{N}_{L}$  & $-5$ & $1$ & $-5$ \\
10  & 5 $(D,\bar{E^{0}},E^{+})_{L}$  & $2$ & $-2$ & $4$ \\
    & $ 5^{*} (\bar{D},E^{0},E^{-})_{L}$  & $-2$ & $-2$ & $1$ \\
1   & $ 1  S^{0}_{L}$  & $0$  & $4$  & $-5$ \\
\hline
\end{tabular}
\end{table}

The ``sequential standard model'' (SSM) boson
$Z''$ has the same couplings as the ordinary $Z$; it cannot
occur in extended gauge theories, but could occur in composite models.
It is   a useful reference point for  comparing the sensitivity of
experimental signals. However, the more realistic cases above
 have weaker couplings to the ordinary fermions.

There are numerous other $Z'$ models. 
In particular, the superstring models based on 
the free fermionic constructions  yield couplings for the 
additional $Z'$'s which do not in general  correspond to the standard
GUT-type models.

We also mention some types of models that have been inspired on
phenomenological grounds.
The leptophobic models, in which the leptons do not
carry any $U(1)'$
charges~\cite{heavylep,lightlep,leptoconstraints},
were motivated by a
reported excess in the $Z \rightarrow b \bar{b}$ branching ratio from
LEP~\cite{lepdata}, which could be accounted for by a small $Z-Z'$ mixing.
Effects in leptonic channels such as $e^+ e^- \rightarrow Z
\rightarrow \mu^+ \mu^-$ would be higher order in the mixing and
therefore negligible. Leptophobic models with both heavy
($\sim$ 1 TeV)~\cite{heavylep} and light (e.g., 150 GeV)
$Z'$~\cite{lightlep} were constructed.
It was shown that a leptophobic variant of the $\eta$ model
could emerge from an $E_6$ 
supersymmetric-GUT with kinetic mixing (see below)~\cite{bkm},
and the possible origin of such models from
superstring theory
was discussed in~\cite{stringylepto}.
Another class of models motivated by $Z \rightarrow b \bar{b}$
involve an extra
$Z'$ that couples only to the third family~\cite{thirdfam}.
However, more recent
LEP and SLD data have significantly
weakened the motivation for both classes of models.
One can also consider fermiophobic models, which do not
couple to ordinary fermions, but which shift the ordinary $Z$ mass
due to mixing~\cite{fermio}.

\subsection{Mass and Kinetic Mixing}

The $Z-Z'$ mass matrix takes the form
\begin{equation}
M^2 = \left( \begin{array}{cc}
	M_Z^2 & \gamma M_Z^2 \\
	\gamma M_Z^2 & M_{Z'}^2
	\end{array} \right) ,
\label{ZZmass}
\end{equation}
where $\gamma$ is determined within each model once the Higgs sector is 
specified.  The physical  (mass) eigenstates of mass $M_{1,2}$
are then
\begin{eqnarray}
Z_1 = + Z \cos\phi + Z' \sin \phi \nonumber \\
Z_2 = - Z \sin\phi + Z' \cos \phi ,
\end{eqnarray}
where $Z_1$ is the known boson
and $\tan 2\phi
=2\gamma M_Z^2/(M_{Z'}^2 -M_{Z}^2)$.
The mass $M_1$ is shifted from the SM value $M_Z$
by the mixing, so that $M_Z^2 - M_1^2 = \tan^2 \phi (M_2^2 - M_Z^2)$.
For $M_Z \ll M_{Z'}$ one has $M_2 \sim M_{Z'}$ and
$\phi \sim \gamma M_1^2/M_2^2$. 

There may also be a gauge invariant
mixing of  the $U(1)$ gauge boson kinetic  
energy terms. This is  equivalent, by appropriate redefinition of the 
 gauge fields, to a mixing between the renormalization group equations
(RGE's) for the running $U(1)$ gauge couplings.
A significant mixing could  be induced by field  
theoretical loop effects~\cite{kmix1,kmix2,ACLIV,bkm}, which occur in
the RGE's when $Tr (Q_Y Q) \ne 0$, with the trace  restricted to
the light degrees of freedom. This can occur in GUTs, for example,
when multiplets are split into light and heavy sectors.
Kinetic mixing can also be
due to higher genus  
effects in superstring theory~\cite{mixdan}. However,  such  
effects are  small for superstring vacua based on the free fermionic
construction~\footnote{If
one of the $U(1)$'s is associated with a large supersymmetry-breaking
$D$-term in a ``hidden'' sector,  the kinetic mixing could propagate
this large scale to the observable sector~\cite{mixdan}.}, 
on the order of at most 
a \%.
An important implication of kinetic mixing is that the effective
charge of the $Z'$ at low energies may contain a component
proportional to the ordinary weak hypercharge~\footnote{The $Z'$
may also contain a component of hypercharge in models
which are not based on $U(1)_Y \times U(1)'$, such as the LR.}.

\subsection{Precision Electroweak and Collider Limits}
Constraints can be placed on the
existence of $Z'$'s either indirectly from fits to high precision
electroweak data
 or from direct searches at  colliders.

\subsubsection{Precision Limits}

In the analyses presented in this and the following  Subsection,
GUT based  $Z'$ models,  described at the beginning of the previous
  Subsection, were
 used.  Although they represent a small subset of models, they are
 representative.

 We list the  constraints on the $Z'$ mass and mixing from an updated
 global fit to precision data~\cite{jepl}
 in Table \ref{cvetictab1}. $M_2$ is mainly constrained by the effects
of $Z_2$ exchange on low energy weak neutral current processes, while
the $Z-Z'$ mixing angle $\phi$ is restricted by the precise
$Z$ pole data from LEP and SLC. The unconstrained fits make no assumption
on the $U(1)'$ charges
of the Higgs doublets which break $SU(2)_L$ and lead to $Z-Z'$ mixing,
so that $M_2$ and $\phi$ are independent.
They lead to relatively weak
limits on $M_2$ but stringent constraints~\footnote{Larger 
(a few per cent) mixings 
are possible in leptophobic and fermiophobic 
models with a light (e.g., 150 GeV) 
$Z'$~\cite{heavylep,lightlep,leptoconstraints,fermio,leptolimits}.
Even with no couplings to leptons, the mixing is constrained by 
the modification of $M_1$ from the SM value.} 
on $|\phi|$,
less than a few times $10^{-3}$. 
The constrained fits assume definite $U(1)'$ charges
for the Higgs doublets (Table~\ref{zprime}). For the $\chi$ and LR
models, the two Higgs doublets, which transform like the $SO(10)$ 10-plet,
have the same $Q Q_{3L}$, so that $\gamma$ in (\ref{ZZmass}) is fixed.
One then obtains a strong constraint on $M_2$ indirectly from the
$\phi$ constraint. For the $\psi$ and $\eta$ models $\gamma$ depends
on the ratio of the two Higgs VEV's, so there is no additional constraint.

\begin{table}[ht]
\caption[]{
Current constraints on $M_2 \sim
M_{Z'}$ (in GeV) for typical models from direct
production at the TEVATRON (${\cal L}_{int}=110$ pb$^{-1}$),
assuming decays into SM particles only,
as well as
indirect limits from a global electroweak analysis (95\%
C.L.). Also shown are the 95\% C.L. limits on the $Z-Z'$ mixing
angle $\phi$.}
\label{cvetictab1}
\centering
\begin{tabular}{|l|ccccc|} \hline
&direct&indirect & indirect & $\phi_{min}$ & $\phi_{max}$ \\ [-3pt]
& & (unconstrained)&(constrained) & & \\
\hline
$\chi$ & 595  & 330 & 1160 & $-0.0029$  &  $+0.0011$ \\
$\psi$ & 590  & 170 & & $-0.0022$  &  $+0.0026$ \\
$\eta$ & 620 & 220 & & $-0.0055$  &  $+0.0021$ \\
$LR$ & 630 & 390 & 1680 & $-0.0013$  &  $+0.0021$ \\
$SSM$ & 690 & 990 &  & $-0.0023$  &  $+0.0004$ \\ \hline
\end{tabular}
\end{table}

There are additional constraints from the absence of
flavor changing processes. Models with $Z'$ couplings 
that are not family universal~\cite{thirdfam} in the
original (weak eigenstate) basis will
generally acquire off-diagonal $Z'$ couplings of the order of the
CKM mixing when the fermions are reexpressed
in the mass eigenstate basis~\cite{nonunivz}.
Even family-universal couplings may acquire off diagonal terms due
to mixing between ordinary and exotic fermions~\cite{london,nardi}.
There will also be loop effects due to the mismatch between
particle and sparticle mixing, leading to off-diagonal gaugino
vertices~\cite{suematsu} analogous to those of the MSSM.

\subsubsection{Collider Limits}

The highest mass limits come from direct searches
at the Tevatron~\cite{colliderlim}.  The most stringent limits,
from CDF, are  obtained by looking
for high invariant mass lepton pairs from Drell-Yan
production of  $Z'$s and their subsequent decay
to lepton pairs,  $p\bar{p}\to Z' \to \ell^+\ell^-$.
The most recent CDF 95\% confidence level results based on ${\cal
L}_{int}=110$ $\hbox{pb}^{-1}$ are listed in Table \ref{cvetictab1}.

The CDF limits in Table \ref{cvetictab1} assume that the
$Z'$ can only decay into the known SM particles.
The limits would be somewhat weaker (by 100--150 GeV) if there
are open channels for the $Z'$ to decay into exotic 
particles~\cite{exoticdecay} (usually expected in such models)
or into new supersymmetric particles, which lower the leptonic branching
ratio.
Whether such decays are kinematically allowed is very model dependent. A 
detailed  analysis of the effects of decays into not only scalar quarks
and leptons, but also neutralinos, charginos, and Higgs bosons has
been carried out for  a class of $E_6$ based  models and  a specific set of MSSM 
parameters
in~\cite{gkk}. Other recent studies have been concerned  with the decays of
leptophobic $Z'$s into exotics~\cite{jlr} or into a scalar
plus $W$ or $Z$~\cite{gg}.

\subsection{Discovery and Diagnostics at Future Colliders}

The expected sensitivity of planned and proposed $p \bar{p}$,
$pp$, and $e^+ e^-$ colliders, including Tevatron upgrades, the LHC,
and various possible NLC scenarios, are listed in Table~\ref{cvetictab2},
taken from~\cite{CG}.
The sensitivities of other proposed machines
are discussed in~\cite{GHP,sg}.
The results for HERA and possible future $ep$ colliders, in which
the $Z'$ are exchanged in the $t$-channel, are not competitive.

\subsubsection{Discovery Limits}

The signal for a $Z'$ at a hadron collider consists of
Drell-Yan production of  high invariant mass lepton pairs.
From  Table \ref{cvetictab2},
 the LHC will be sensitive
to reference $Z'$s up to several TeV, well above the expected
 maximum of 1 TeV for a class of superstring-motivated models discussed in Section 4.
The discovery limits
are relatively insensitive to
specific GUT models and are   fairly robust.  

At $e^+e^-$ colliders discovery limits are indirect,  inferred
from deviations from the SM predictions for various
cross sections and asymmetries, due to
interference between the $Z'$, $\gamma$, and $Z^0$
propagators~\cite{e+e-}.
The basic process is $e^+e^- \to f\bar{f}$, where $f$
could be leptons $(e,\; \mu ,\; \tau)$ or quarks $(u, \; d, \; c,\;
s,\; b, \; t)$. The  observables include: 
({i}) The leptonic cross
section, $\sigma^\mu = \sigma (e^+ e^- \to \mu^+ \mu^-)$.
({ii}) The ratio of the hadronic to the QED point cross section,
$R^{had}= \sigma^{had}/\sigma_0$.
({iii}) The leptonic forward-backward
asymmetry, $A^\ell_{FB}$,  and if $c$ and $b$
quark flavor tagging were sufficiently efficient, $A^c_{FB}$ and $A^b_{FB}$.
({iv}) The leptonic and hadronic  polarization asymmetries,
$A^\ell_{LR}$,
$A^{had}_{LR}$, and final state
polarization of $\tau$'s,
$A_{pol}^\tau$.
({v}) The polarized forward-backward asymmetry for
specific fermion flavors, $A^f_{FB}(pol)$.

From Table \ref{cvetictab2}
 the NLC sensitivity
is somewhat less than that of the LHC
for $\sqrt{s} =$ 500 GeV, but increases rapidly for
higher energies.
However, the bounds  are more model dependent. 
Also, since the bounds are indirect, they are  
especially sensitive to systematic and theoretical uncertainties.

\begin{table}[ht]
\caption[]{
Bounds on $M_{Z'}$ (in GeV)
for typical models achievable at
proposed hadron and $e^+e^-$ colliders.
The discovery limits  for $Z'$
at hadron colliders  are
for typical models  with  10 events in   $e^+e^-\ +\ \mu^+\mu^-$,
while those for
$e^+e^-$ colliders are at 99\% C.L. obtained from
$\sigma^\mu$, $R^{had}$, $A_{LR}^{\mu}$, and
$A^{had}_{LR}$. The   table is from~\cite{CG}. }
\label{cvetictab2}
\centering
\small
\begin{tabular}{|l|cc|cccc|}
\hline
Collider &$\sqrt{s}$ [TeV] &${\cal L}_{int}\  [\hbox {fb}^{-1}]$& $\chi$
& $\psi$ & $\eta$ & $LR$  \\
\hline
TEVATRON ($p\bar{p})$ &
 \91.8& \9\91&  \9775 & \9775 & \9795 & \9825 \\
TEVATRON$^\prime$ ($p\bar{p})$ &
 \92\9\ & \910&  1040 & 1050 & 1070 & 1100  \\
LHC ($pp)$ &
  10\9\ & \940&  3040 & 2910 & 2980 & 3150  \\
LHC ($pp)$ &
  14\9\ & 100 &  4380 & 4190 & 4290 & 4530 \\
\hline
LEP200 ($e^+e^-$) &
  0.2& \9\90.5& \9695 & \9269 & \9431 & \9493   \\
NLC ($e^+e^-$) &
  0.5&\9  50\9\ &  3340 & \9978 & 1990 & 2560   \\
NLC-A ($e^+e^-$) &
  1.0 & 200\9\ & 6670 &  1940 & 3980 & 5090  \\
NLC-B ($e^+e^-$) &
  1.5 & 200\9\ & 8220 &  2550 & 4970 & 6240   \\
NLC-C ($e^+e^-$) &
  2.0 & 200\9\ & 9560 &  3150 & 5830 & 7210   \\
\hline
\end{tabular}
\end{table}

\subsubsection{ $Z'$ Diagnostics at the LHC and NLC }

Following the discovery of a heavy $Z'$, one would want to determine
its chiral couplings to quarks and leptons.
Their ratios would allow one to discriminate
between models and to determine
the nature of the underlying  extended gauge structure~\cite{ACLIV}.

 For masses up to $\sim$ 1--2 TeV considerable information 
should be obtainable from the LHC and NLC, with
their capabilities complementary. The range $M_{Z'} < $ 1 TeV,
expected in a class of  models  discussed in Section 4, 
should allow precise determination of a
number of useful diagnostics.


 In the main LHC production channels,  $pp \to Z'\to \ell^+
\ell^-$ ($\ell=e,\mu$), one would be able to measure immediately
the mass $M_{Z'}$, the width $\Gamma_{tot}$ and the leptonic cross section
 $\sigma_{\ell \ell}$. By itself, $\sigma_{\ell\ell}=\sigma (pp\to Z')
 B$  is {\it not}   a useful diagnostic
probe for the $Z'$ gauge couplings to quarks and leptons:
while $\sigma(pp\to Z')$   can be
calculated to within  a few percent for
given $Z'$ couplings, the branching ratio into leptons,
$B\equiv\Gamma(Z'\to\ell^+\ell^-)/\Gamma_{tot}$,
depends on the contribution of exotic
fermions and supersymmetric partners to the
$Z'$ width.
However, $\sigma_{\ell \ell}$
would be a useful indirect probe for the existence of the exotic
fermions or superpartners.
On the other hand, the product
$\sigma_{\ell \ell} \Gamma_{tot} = 
\sigma \Gamma(Z'\rightarrow \ell^+\ell^-)$ does
 probe the absolute magnitude of the gauge couplings.

The most useful diagnostics are signals  which probe {\it  relative  
strengths} of $Z'$ gauge
couplings to ordinary quarks and leptons. The  forward-backward asymmetry (as a function of the $Z'$
rapidity) in
the main ($\ell^+\ell^-$) production channel
was the first recognized  and most powerful probe for the  gauge couplings
 at future hadron  colliders~\cite{LRR}.
Since then a number of new,  complementary  probes were
proposed (for details, see~\cite{CG}).
The nature of such probes can be classified according to the
type of  channel in which they can be measured:
({i}) In the main production channels one can measure the
forward-backward asymmetry~\cite{LRR,dittmar}, the 
ratio of cross-sections in different rapidity bins~\cite{ACL},
and asymmetries if proton polarization  were available~\cite{FT}.
({ii}) Possible observables in other two-fermion final state channels
include the polarization of produced $\tau$'s~\cite{AAC}
and the $pp\to Z'\to jet\  jet$ cross section~\cite{RM,MOH}.
({iii})  In four-fermion final state  channels
one may observe the rates for 
rare decays $Z'\to W\ell\nu_{\ell}$~\cite{RI,CLII}
and for associated productions $pp\rightarrow Z' V$
    with $V=(Z,W)$~\cite{CLIV}  and $V=\gamma$~\cite{RII}.

 At  the LHC the above signals are feasible
 diagnostics  for $M_{Z'}\lsim 1-2$ TeV, with
 large
luminosity being important.
For higher $Z'$ masses the number of events drops rapidly,
and by $M_{Z'}\sim 3$ TeV there is little ability
to distinguish between models.

Similarly, the indirect probes described above for virtual $Z'$ at
the NLC allow for  diagnostics of the $Z'$ couplings
for $M_{Z'}\lsim 1$ TeV~\cite{e+e-,ACLIII,HREE,ACLIV},
especially if polarization and efficient
tagging of heavy flavors $(c,b,t)$ are available. However, by
$M_{Z'}\sim $ 2 TeV the uncertainties for the couplings in the
typical models are too large to discriminate between models.

The LHC and NLC are complementary~\cite{ACLIII}.
For the projected luminosities, both
should be able to determine various ratios of $Z'$ couplings reasonably
well for sufficiently small $M_{Z'}$. 
For a 1 TeV $Z'$ the LHC error bars on the extracted 
quantities are smaller by a factor $\sim$ 2 than those at the NLC,
but there are typically sign ambiguities for the $U(1)'$ charges of quarks and
leptons  that can be resolved at
the NLC.
Thus,  the LHC and the NLC together
have the potential to uniquely determine
$M_{Z'}$, an overall $Z'$ gauge coupling  
strength, as well
as a  unique determination of   { all}  (four)  quark and lepton charge
ratios with
error bars in the 10--20\% range, provided $M_{Z'}\lsim 1$--2 TeV.


\subsection{Constraints on Exotic Fermions}

 There have also been studies of the present and future
constraints on possible exotic particles~\cite{exotic}.
For example, some models predict the existence of a heavy
vector ($SU(2)_L$-singlet) charge $-1/3$ quark, $D_L-D_R$, which could
be produced at a hadron collider by ordinary QCD processes and
decay by $D_L-b_L,s_L,d_L$ mixing into, {\it e.g.,} $cW$, $bZ$,
or $bH$,
where $H$ is a neutral Higgs boson. 
Precision experiments (weak charged current, neutral
current, and flavor changing constaints) typically imply that the
mixings between $D_L$ and $d_L$ is less than $\sim0.01$.
Typically, the $cW$,
$bZ$, and $bH$ decays occur in the ratio $\simeq 2:1:1$. Currently,
$m_D > $ 85 GeV if it mixes mainly with $b$~\cite{exotic}.
Heavy gauge bosons and exotic matter constraints
have been addressed together in~\cite{nrl}.

\section{$Z'$s--Theoretical Considerations}
$Z'$ models fall into different classes, depending on whether  they  are
embedded into a  GUT  and whether
supersymmetry is included. In the
following we briefly summarize features of different types of models.

\subsubsection{$Z'$ Models  in GUT's  without  Supersymmetry}

In a  general class of models with extended gauge structure
which  do not incorporate constraints
of 
supersymmetry, supergravity or superstring theory,
 the mass  and couplings 
of additional gauge bosons are 
free parameters in general. However,  one  
class of  models of special interest is  based on  the  GUT 
gauge structure~\footnote{For a  review, see~\cite{Langacker} 
and references therein.}. At $M_U$  the GUT  gauge group is
 broken  to a smaller one
which includes the SM gauge group and  may also  include  additional $U(1)'$
factors. As opposed to  general models 
 the gauge couplings of the additional $Z'$ are  now determined
within each 
 GUT model, and the quantum numbers of additional exotic
particles are also  fixed (see  Section 2, Table 1, for  examples).
 However, even within  
GUT models, 
there is {\it no  robust prediction for the $Z'$ mass and the 
mass of the accompanying 
exotic particles};
without fine-tuning of parameters their masses are likely to be at
$M_U$, while with fine-tuning  their masses can be anywhere
between $M_U$ and $M_Z$.

\subsubsection{$Z'$ Models in  Supersymmetric GUT's} 

Supersymmetric GUT models possess the  advantages  of the
ordinary GUT models,  e.g., gauge coupling unification,
and they may provide more constraints
on the parameters of the theory.
The  GUT models  with  the  
MSSM below $M_U$ have a gauge coupling unification~\cite{SUSYGUTS}
 that is consistent with current
experiments. 
Taking the observed $\alpha$ and weak
 angle $\sin^2 \theta_W$ as inputs
 and extrapolating assuming the particle content 
of the MSSM, one finds~\cite{LP}
that the running $SU(2)_L$ and $U(1)_Y$  couplings meet at a
scale $M_U \sim 3 \times 10^{16}$ GeV. One can then predict
$\alpha_s(M_Z) \sim 0.130 \pm 0.010$ for the strong coupling. The
actual value of $\alpha_s(M_Z)$ is still controversial, with determinations
generally in the range 0.11--0.125~\cite{ALS}, but is
roughly   within the MSSM 
prediction.  To ensure correct electroweak symmetry
 breaking,  fine-tuning of the superpotential parameters  
 or a specific (higher-dimensional) Higgs
 representation, e.g., the ``missing partner'' mechanism~\cite{DW,Nilles}, 
 is needed.
 
On the other hand, within   a symmetry breaking scheme  in which 
  the GUT group is broken down  to the SM  and additional $U(1)'$
 factors, the success of gauge coupling unification 
  can be ensured only
for certain specific exotic particle spectra.
This in general  involves complicated 
 fine-tuning of certain
  parameters in the superpotential. (One exception is a special case
  with   complete (light) GUT
  multiplets contributing to the  renormalization group flow.)  
 In this case,  the  parameters 
  of the superpotential should  be further constrained
  to ensure additional  symmetry breaking.
  In particular, the  breaking of $U(1)'$  may involve large  Higgs
  representations and/or fine-tuning 
to ensure $M_{Z'}\ll M_U$ and 
  $D$-  and $F$-flatness up to ${\cal O}(M_{Z'})$. 
  Even further
  fine-tuning  of  the  superpotential parameters would be needed 
   to  allow
     $M_{Z'}\gg M_Z$~\footnote{A somewhat
   analogous analysis of the
  symmetry breaking pattern of supersymmetric $SO(10)$    down to the
  left-right symmetric models, and then further breaking  to the SM gauge
  group,  was  addressed in~\cite{AM}.}. 
  A study of (leptophobic) $Z'$ within supersymmetric  $E_6$ 
   was recently addressed
  in~\cite{bkm}.
  
Thus, in spite of the  constraining structure  
of supersymmetric GUT models,  the  mass of $Z'$ and the accompanying exotic
particles  again
generically   tends to be of
order $M_U$.    Other mass scales 
 could be achieved by choosing specific  Higgs
  chiral supermultiplets  and  adjusting the amount  of fine-tuning for the
superpotential parameters. 

\subsubsection{Supersymmetric $Z'$  Models without GUT Embedding} 

Supersymmetric  models  with  additional $U(1)'$ factors, 
but without GUT embedding, may provide a promising 
class of models with
  more predictive power. In a certain way they provide a {\it minimal
  extension of the MSSM}.  In the case in which  mass parameters are absent in
  the superpotential, the  soft supersymmetry breaking mass parameters and the 
  type of the SM singlets responsible for the $U(1)'$ breaking determine
  the mass scale of the $Z'$ without  additional
   (excessive) fine-tuning of parameters.

  Such a class of models can be derived within  certain 
   classes of superstring compactifications. 
   In these models
the massless particle spectrum  and  couplings in the superpotential are
calculable,  and there are 
 {\it no mass parameters in the superpotential}.  Thus,
  supersymmetric $Z'$ models with built in constraints on
   couplings from   superstring models 
    {\it provide a class of models with  a predictive power}
      for  $Z'$ physics and  that of the accompanying exotic particles. 
  
In the following  Section we
 review the  properties of $Z'$s based on that type of model.  
It is primarily based on  Refs.~\cite{CL,CDEEL,CCEEL}. 
An important related analysis 
was given in~\cite{SY}~\footnote{Some aspects of  $Z'$s in superstring 
models were also
addressed in~\cite{stringylepto}.}.  
Earlier  work,  which addresses $Z'$s in  models
 with softly broken $N=1$
 supergravity  with  no direct connection to superstring models, but 
 with $Z'$ quantum
 numbers obtained from  $E_6$  embeddings (i.e., with $E_6$ 
 broken already
  at the string scale by the Wilson line (Hosotani) mechanism), 
  was given in~\cite{GRH}.  More recent analyses in this context
  appeared in~\cite{keithma,gkk}.

\section{$Z'$s from  Superstring Compactifications}


A class of superstring compactifications that are based on  fermionic 
 constructions~\cite{ABK,NAHE,faraggi90a,CHL,AFIU,Kakushadze,dienes} 
  provides the necessary ingredients for radiative $U(1)'$ breaking, either
  at the electroweak or at an intermediate scale.
The particle content and the gauge group 
 structure contain, along with the  MSSM, 
  additional $U(1)'$ gauge symmetry factors. 
  The massless particle spectrum and the
  superpotential couplings (which do not have mass terms) are calculable.
  Importantly, 
  in this class of superstring models 
the  trilinear  (Yukawa) couplings in the superpotential 
are  either absent or  of order one.

In the following Subsections we will summarize 
 the general properties of such
superstring vacua and then concentrate on the two types of $U(1)'$
symmetry breaking pattern.

\subsection{Properties of Superstring Vacua}

Weakly coupled  heterotic string theory, compactified down to 
 $N=1$ supersymmetric theory in four-dimensions, 
  possesses a number of attractive features, such as
    a gauge group  that contains the SM 
    gauge group and  massless chiral supermultiplets, 
    which could  play the role of family  and Higgs supermultiplets. 

The   theory also  predicts  (at tree level)
 gauge coupling unification at 
$\MS$ $\sim $ $g_U \times 5 \times 10^{17} $ GeV~\cite{K},
 where $g_U$ is the gauge coupling at the string scale. 
This is to be compared with the MSSM value 
$M_U \sim 3 \times 10^{16}$ GeV. When  properly viewed as predictions 
 for $\ln(M_U/M_Z)$
and $1/\alpha_s$, the gauge unification works to within
10--15\%~\footnote{This discrepancy can be remedied~\cite{BL}, e.g.,  
by introducing  exotic particles with 
specific mass ranges. One must, however, be careful because generic exotics
can introduce corrections of ${\cal O}(1)$. Another possiblity is 
to   embed  $U(1)_Y$ with $k<5/3$. }.

Using the powerful duality symmetries in superstring 
 theory one can obtain~\cite{WittenII} a handle on certain properties 
 of the strongly coupled heterotic string as well. In particular,
  for the strongly coupled  $SO(10)$ heterotic string theory,  
  which is dual to a weakly coupled Type I superstring theory, $\MS$,  the 
  scale of the four-dimensional gauge coupling unification,   depends on
 the  value  of  the ten-dimensional dilaton of the Type I superstring theory. 
 $\MS$  can  be made compatible with $M_U$
 for a small enough value of the dilaton (weakly coupled Type I superstring
 theory). 
 Similarly, for the strongly coupled $E_8\times E_8$ heterotic
  string theory, which  is  dual to  eleven dimensional $M$-theory 
compactified on
  $S^1/Z_2$ (one-dimensional $Z_2$ orbifold), 
   $\MS$ 
  depends  on  the radius of the circle $S^1$. $\MS$ can be made compatible
  with $M_U$ for the values of the radius
   which  correspond to the
  limit in which one of the $E_8$'s  is about to become strongly
  coupled~\cite{WittenII}. 
 Unfortunately,  
  explicit constructions of such superstring 
   vacua  and their  phenomenological analyses are still in their infancy.

There are two  major obstacles to connecting 
the string vacua of the weakly  coupled heterotic string to 
 the low energy   world: (i) 
The absence of a fully satisfactory
 scenario for supersymmetry 
  breaking, either at the level
 of world-sheet dynamics or  of the effective field theory. 
The supersymmetry breaking is expected to induce soft mass parameters which
provide another scale in the theory that can hopefully 
provide a link between  $\MS$ and  the 
electroweak scale.
(ii) 
There is a large degeneracy of superstring vacua, i.e., there 
are by now a large number of  superstring models, all dynamically on the same 
footing. 

 One may parametrize  our ignorance  
of supersymmetry breaking
 by  introducing  soft  supersymmetry breaking  mass terms. 
 In this manner one looses definitive predictive power, i.e., 
 the quantitative results  critically depend on the values of 
 such parameters. However,  the explicitly calculable structure
of
the superpotential (and other terms in the supersymmetric part 
of the effective Lagrangian),  still allows for  useful
predictions for the  low energy physics.  

As for the issue of degeneracy of superstring vacua,
the  GUT models based on higher level Ka\v c-Moody 
algebra constructions~\cite{CHL,AFIU,Kakushadze} 
 provide  the needed adjoint representations; however,  they possess large
additional gauge group factors,  and a large number of 
 additional  exotic particles.  
It may therefore be advantageous to study  a class of models which
     at $\MS$ already possess  the SM  gauge group as a part of the
observable sector gauge structure, and  have a massless particle 
content that includes 
 three SM families  and  at least two SM
Higgs doublets.
A number of such superstring models 
(not necessarily consistent with gauge coupling unification)
were constructed. One class~\cite{NAHE,faraggi90a,CHL} is 
based on the free    fermionic constructions. 
 A set of such  models   also constitutes a starting point to address 
 phenomenological issues of $Z'$  physics 
in supersymmetric models based on superstring compactifications.

\subsection{Anomalous and Nonanomalous $U(1)'$ in Superstring Theory}

A class of models  based on  free fermionic constructions
  have a number of  generic features. They are $N=1$ supersymmetric
    string models with  
the  SM gauge group  $SU(2)_L\times U(1)_Y\times SU(3)_C$,  
three ordinary families,  and at least two SM
doublets. Such models  in general also contain   non-Abelian 
``shadow'' sector  gauge symmetries  and {\it a  number of 
 additional $U(1)'$'s, one of them 
generically anomalous}.  
The shadow sector non-Abelian gauge group  may play a role in 
dynamical supersymmetry breaking~\cite{dudas}.
In addition, there are  a large number of additional matter multiplets,
which  transform non-trivially under the $U(1)'$  and/or the  standard 
model  symmetry~\footnote{We  select only
those models with  $SU(3)_C$, $SU(2)_L$ and $U(1)_Y$  
 embedded into  $SU(5)$, since for
other types of embedding  the normalization of  the $U(1)_Y$ gauge 
coupling is different from the one leading to 
gauge coupling unification in the MSSM.
The fact that the observable sector gauge 
group at $M_{string}$ is  not $SU(5)$ but the SM gauge group 
implies~\cite{S} that the theory  contains 
fractionally charged color singlets.}. 

Due to the  anomalous  $U(1)'$  symmetry, at  genus-one there 
is an additional contribution  of  ${\cal O}(\MS^2/16\pi^2)$~\cite{DSW}
 to the corresponding $D$-term in the effective Lagrangian.  The 
  numerical suppression factor of ${\cal O} (1/16\pi^2)$ renders 
  the scale of the genus-one contribution to be generically
  smaller than $M_{string}$
  by  one to two orders of magnitude. The contribution of 
  such a term is canceled~\cite{DKI,DSW} by 
 giving nonzero  vacuum 
  expectation values (VEV's)   of ${\cal O}(M_{string}/10)$ to certain 
  multiplets in such a way  that the  $D$-flatness and  $F$-flatness 
  condition is maintained at genus-one level of the  effective string theory,
  thus providing a  mechanism for  ``restabilizing'' the 
vacuum~\footnote{ 
The SM singlets which break the anomalous $U(1)'$   can  couple through  
appropriate
nonrenormalizable terms to the SM  particles, introducing
 an effective $\mu$-parameter~\cite{shrock} and
 appropriate fermionic masses~\cite{F,binetruy96}.    
Since the VEV of these singlets is
${\cal O}(\MS/10)$ the needed non-renormalizable terms are usually 
of  high order.}.
The nonzero VEV's typically
 break  a number of additional
   non-anomalous $U(1)'$'s 
   at  $\MS$  as well.  Thus, the  gauge symmetry and the   
light exotic particle content of  the observable sector
is in general drastically reduced. Nevertheless,
there  often remain  one or more {\it non-anomalous $U(1)'$'s } 
and associated  light exotic
matter~\footnote{The study of some  $D$-flatness and $F$-flatness
 conditions within some models  was given, for example,
 in~\cite{faraggi90a,CHL}. 
However a systematic
  study of these conditions for a large class of such models is 
 needed to classify all  possible symmetry breaking
 scenarios.}.

In the following  we shall concentrate on  phenomenological 
consequences  of an additional non-anomalous 
$U(1)'$ symmetry. Potential  additional
 phenomenological problems~\footnote{
For example: (i)  there  may  be additional color triplets  
which could   mediate a too fast 
proton decay;
(ii) the   mass spectrum  
of the ordinary   fermions   may not be  realistic;
and (iii)  the light exotic particle spectrum  may not be consistent with  
 gauge coupling unification.}  will not 
 be addressed.

\subsection{$U(1)'$ Symmetry Breaking Scenarios}

For the sake of simplicity we consider only one 
additional $U(1)'$.  The symmetry breaking  of the additional $U(1)'$ 
 must take place 
via the Higgs mechanism, in which the  scalar components  of 
 chiral supermultiplets $S_i$ which carry  non-zero charges 
 under the $U(1)'$ acquire   non-zero VEV's
and spontaneously break the symmetry. 
 The low energy effective action, responsible for 
  symmetry breaking, is  specified by the
  superpotential, K\"ahler potential, $D$-terms, and soft supersymmetry 
breaking terms. 

There are  two  symmetry breaking scenarios:

\begin{itemize}

\item {\it Electroweak Scale Breaking}

When  an additional $U(1)'$  is broken by a  {\it single} SM  singlet $S$,
the mass scale of the $U(1)'$ breaking should 
be~\cite{CL} in the electroweak range (and not
larger than a TeV).  
The $U(1)'$ breaking may be radiative due to the  large (of order one) 
Yukawa couplings of  $S$ to exotic particles.  The scale
of the symmetry breaking is set by the soft supersymmetry breaking
scale, in analogy to the radiative breaking of the electroweak
symmetry  due to the large top Yukawa coupling.
The  analysis was generalized~\cite{CDEEL,SY} 
by  including the coupling of  the two SM Higgs
doublets to the SM singlet $S$.
This class of models not only  predicts the
existence of additional gauge bosons and exotic matter particles, but can
ensure that their masses are in the electroweak range.
Depending on the values of the assumed soft supersymmetry breaking
mass parameters at $\MS$, each specific model leads to calculable
predictions, which can satisfy the phenomenological bounds.  In addition,
the models 
forbid an elementary $\mu$-parameter for
appropriate $U(1)'$ charges of the chiral superfields,
 but an effective $\mu$ is generated by the
electroweak scale VEV of the singlet, thus providing a natural solution to
the $\mu$ problem.

\item {\it Intermediate Scale Breaking}

The second scenario takes place~\cite{CL} if there are couplings in the
renormalizable superpotential of exotic particles to two or more
mirror-like singlets $S_i$ with opposite  signs of their $U(1)'$
 charges. In this case, the
potential may have $D$- and $F$-flat directions, along which it consists only of
the quadratic soft supersymmetry breaking  mass
 terms.  If there is a mechanism to drive a  particular linear combination 
negative at
$\mu_{ RAD}\gg M_{Z}$, the $U(1)'$ breaking is at {\it an intermediate scale}
of order $\mu_{RAD}$, or in the case of  dominant non-renormalizable terms in the
superpotential  the $U(1)'$ takes place at an intermediate scale governed by these
terms. 

\end{itemize}

In the following we shall summarize the features of the two scenarios. 
For more details see~\cite{CDEEL} and~\cite{CCEEL}, respectively.

\subsection{Electroweak Scale Breaking}
    
The superpotential in this  model is  of the form:
\begin{eqnarray} 
W=h_{s}\hat{S}\hat{H}_{1}\cdot\hat{H}_{2} + h_Q \hat{U}_3^c\hat{Q}_3\cdot\hat{H}_2, 
\label{superpot}
\end{eqnarray}
where
the  Higgs doublet  superfield $\hat{H}_2$  has only a  Yukawa coupling to a
single (third) quark family~\footnote{The masses of  other  quarks and leptons
are obtained in this class of models through
non-renormalizable  terms~\cite{F}. The 
fermion masses  obtained in this way 
may not be realistic.}. 
For simplicity,
 the 
K\" ahler potential is written in a canonical form, providing
 canonical kinetic energy terms for the matter superfields.
Supersymmetry breaking  is parametrized with
the most general
soft mass parameters, with $m_{1,2,S,Q,U}^2$ corresponding to the soft 
mass-squared terms of the scalar fields  $H_1,\ H_2, \ S ,\   Q_3, \ U_3^c$,
respectively, and   $A$ and $A_Q$  are  
the soft mass parameters 
associated with the
first and the second trilinear terms in the superpotential (\ref{superpot}).

Gauge symmetry breaking is now driven by the VEV's
of the doublets $H_1$, $H_2$ and the singlet $S$,  obtained 
by  minimizing  the Higgs potential.
By an appropriate choice of the global phases of the fields,
 one can choose  the VEV's  
 such that $\langle H_{1,2}^0\rangle=v_{1,2}/\sqrt{2}$  
 and $\langle S\rangle =s/\sqrt2$ are 
positive in the true minimum.
Whether the obtained  local minimum of the potential
is  acceptable will depend on the location and 
depth of the other possible minima and of the barrier height and width 
between the minima~\cite{KLS}.

A nonzero value (in the electroweak range) for $s$ 
 renders   the first term of the superpotential
(\ref{superpot}) into an effective $\mu$-parameter, i.e., 
$\mu {\hat H}_1 {\hat H}_2$, with 
 $\mu\equiv h_ss/\sqrt2$  in  the electroweak range, thus
  providing an elegant solution to the $\mu$ problem.
  
The $Z-Z'$ mass-squared
matrix is given by  (\ref{ZZmass}), 
where 
\begin{eqnarray}
M_{Z}^{2}&=&\frac{1}{4}G^2(v_{1}^2+v_{2}^2),\\
M_{Z'}^{2}&=&{g'}^{2}(v_{1}^{2}Q_{1}^{2}+v_{2}^{2}Q_{2}^{2}+s^{2}Q_{S}^{2}),\\
\label{mix}
\gamma&=&{2g'\over{G}}{{{(v_{1}^2Q_{1}-v_{2}^2Q_{2})}}\over{v_1^2+v_2^2}}.
\end{eqnarray}
Here $g'$  is the gauge coupling for $U(1)'$,  $G=\sqrt{g_Y^2+g'^2}$, 
and $Q_{1,2}$ and $Q_S$ are the $U(1)'$ charges for the
 ${\hat H_{1,2}}$ and
${\hat S}$ superfields.

\subsubsection{Electroweak Scale Conditions}

This symmetry breaking scenario can be classified
 in three different categories 
according to the value of the singlet $s$ VEV:

\begin{itemize}
\item  $s=0$.

In this case  the breaking is driven only by the two Higgs 
doublets
 (this would be the typical case if the soft mass of the singlet remains 
positive). The $Z'$ boson would generically  acquire mass of the same order 
as the $Z$, and 
some particles (Higgses, charginos and neutralinos) would tend to be 
dangerously light~\footnote{In the 
$Q_1+Q_2=0$ (which allows an elementary $\mu$-parameter) or large
$\tan\beta=v_2/v_1\gg 1$  cases  one of the neutral 
gauge bosons becomes  massless~\cite{Luo}.
 This does not provide a
viable hierarchy since the $W^\pm$ mass
is  non-zero and  related to
$v_{1,2}$ and
$G$ in the usual way.}. There is  a possibility of a
small $Z-Z'$ mixing due to  cancellations, and
by considerable fine-tuning one  may be able to 
arrange the parameters to barely satisfy experimental
 constraints.

\item  $s\sim v_{1,2}$.

This case would naturally give $M_{Z'}\simgt M_Z$ (if $g'Q_{1,2}$ is not too small) 
and the   effective $\mu$-parameter  may be small. 
Thus, some sparticles may be expected to be light.  
One requires $Q_1=Q_2$ to have negligible $Z-Z'$ mixing~\footnote{
 Such models are allowed
for, e.g.,  leptophobic
couplings~\cite{heavylep,lightlep,leptoconstraints,bkm,leptolimits}.}.  
A particularly interesting example is the   ``Large
Trilinear Coupling  Scenario'', in which a
 large  trilinear  soft supersymmetry breaking  term  dominates the symmetry
 breaking pattern, and 
relative signs and the magnitudes of the 
soft mass-squared terms are not important since they contribute negligibly to 
the location of the minimum.  The three Higgs fields  assume  
approximately equal VEV's: 
$v_1\sim v_2 \sim s \sim 174 $ GeV.
In this scenario, the electroweak phase
transition may be first order, with
potentially interesting cosmological implications.

\item  $s\gg v_{1,2}$.

Unless $g'Q_S$ is large, $M_{Z'}\gg M_Z$ requires $s \gg v_{1,2}$
and  the effective $\mu$-parameter 
is  naturally large.
In this case  the breaking of the $U(1)'$ is triggered
effectively by the running of
the soft mass-squared term  $m_S^2$ towards negative values in the
infrared, yielding
$s^2\simeq {- (2 m_S^2)}/{({g'}^2Q_S^2)}$ and
$M_{Z'}^2\sim -2 m_S^2$, with $m_S^2$ evaluated at $s$.
The presence of this large singlet VEV influences
$SU(2)_L\times U(1)_Y$ breaking already at tree level.
The hierarchy $M_{Z'}\gg M_{Z}$
 results from a cancellation of different
 mass terms of order
 $M_{Z'}$; the fine-tuning involved is  roughly given by
 $M_{Z'}/M_Z$. The $Z-Z'$ mixing is suppressed by the large $Z'$ mass
 (in addition to any accidental cancellation for particular choices of
 charges).
 Excessive fine-tuning
 would be
 needed for $M_{Z'}\gg 1$ TeV. 
 More details are given in~\cite{CDEEL} (see also~\cite{SY,gkk}).
\end{itemize}

\subsubsection{String Scale Conditions}

We now turn to the renormalization group analysis 
to determine what boundary conditions 
at $\MS$ are required 
to reach the desired low energy parameter space.
The numerical (and some semi-analytic) 
analysis of the RGE's 
is given in~\cite{CDEEL} (see also~\cite{CL,SY,keithma}).
 Here we summarize the results.

The RGE's are solved  numerically as a
function of the boundary conditions at  $\MS$. The 
initial values of the Yukawa couplings (of the
Higgs fields to the singlet and of the Higgs field to
the third quark family) are chosen to be of  the order of magnitude of the gauge
coupling, i.e., $h_s=h_Q={\sqrt 2} g_U$ at $\MS$, as 
determined in a class of superstring models. 
 These couplings provide a dominant
contribution to the RGE's of the soft mass parameters.  

With the minimal particle content, i.e., the MSSM particle content and one SM
 singlet
with couplings specified by the renormalizable part of the superpotential
 (\ref{superpot}), universal soft
supersymmetry breaking mass parameters at $\MS$ do not yield
 phenomenologically acceptable
parameters at the electroweak
scale.  
Acceptable parameters  can be obtained with either:


\begin{itemize}
\item {\it Nonuniversal boundary conditions}


 With the minimal particle content,
   nonuniversal soft
supersymmetry breaking is required at $\MS$
to obtain  the viable gauge symmetry breaking scenarios previously
described~\footnote{The soft mass-squared terms at $\MS$ are assumed to be
positive.}.
In most cases, the gaugino
masses   at  $\MS$ must be  chosen
small relative to the other soft  mass parameters.
For the large trilinear coupling scenario,
the soft mass-squared parameters at $\MS$ are about a factor
of ten larger than their values at the electroweak scale. 
 See~\cite{CDEEL} for
more details.

\item {\it Additional exotics}

Many superstring models predict the existence of  additional exotic particles, 
e.g., 
additional $SU(3)_C$ triplets ${\hat D}_{1,2}$, 
which couple to $\hat S$ via the superpotential:
\begin{equation}
W_D=h_D{\hat S} {\hat D}_1 {\hat D}_2.
\label{sptrip}
\end{equation}
 For 
Yukawa coupling(s)  of the order of the gauge couplings, e.g.,  $h_D\sim
{\sqrt2}g_U$, 
  such exotic particles modify the RGE analysis
significantly~\footnote{Since such exotics may destroy the
gauge coupling 
unification,  one may need additional exotics 
to compensate.}.
A large  singlet VEV can be
obtained when additional color
triplets are present, even with universal
boundary conditions.  
This  was exhibited 
in~\cite{CL}
 in the limit of small
gaugino masses 
and trilinear couplings,  using
semi-analytic expressions for the RGE's (see also~\cite{CDEEL,keithma}).
In contrast, the large trilinear
coupling scenario is more difficult to obtain
 with additional exotic particles, and   
  nonuniversal boundary
conditions are required.

\end{itemize}

\subsubsection{The Spectra of Other Particles}

 The spectrum of physical Higgses after symmetry breaking consists of three 
neutral CP even scalars ($h^0_i$, $i=1,2,3$), one CP odd pseudoscalar ($A^0$) and 
a pair of charged Higgses ($H^\pm$),  i.e., it has one scalar more than in the 
MSSM (for a detailed analysis see~\cite{CDEEL},
for an earlier discussion see~\cite{dterms,GRH}). 
Masses for the three neutral scalars are obtained by diagonalizing the
corresponding $3\times 3$ mass matrix. 
The tree level mass of the lightest scalar $h_1^0$ satisfies the bound
\begin{equation}
\label{bound}
m_{h_1^0}^2\leq M_Z^2\cos^22\beta+\frac{1}{2}h_s^2v^2\sin^2 
2\beta+{g'}^2{\overline Q}_H^2v^2,
\end{equation}
where ${\overline Q}_H=\cos^2\beta Q_1+\sin^2\beta Q_2$. Again $v^2\equiv
v_1^2+v_2^2$ and $\tan\beta\equiv v_2/v_1$. 
 In contrast to the MSSM, $h_1^0$ can be 
heavier than $M_Z$ at tree level, indicating
 that $h_1^0$ can  escape detection at LEPII. 

The Higgs spectrum is 
particularly simple in the large $s$ case.
 The mass of the lightest Higgs boson
$h_1^0$
 remains
below 
the bound (\ref{bound}) and approaches 
\begin{equation}
\label{asympto}
m_{h_1^0}^2\simlt M_Z^2\cos^22\beta+
h_s^2v^2\left[\frac{1}{2}\sin^2 
2\beta-\frac{h_s^2}{{g'}^2Q_S^2}-2\frac{{\overline Q}_H}{Q_S}\right].
\end{equation}
The limiting value (\ref{asympto}) for  $m_{h_1^0}$  can be bigger or smaller 
than the MSSM upper bound 
$M_Z^2\cos^22\beta$, depending on couplings and charge assignments.

The pseudoscalar 
$A^0$ mass $m_{A^0}^2\simeq\sqrt{2}Ah_ss/\sin 2\beta$ is  expected to be 
large (unless $Ah_s$ is very small), and  one of the neutral scalars
and the charged Higgs are then approximately degenerate with $A^0$, completing a
full
$SU(2)_L$ doublet $(H^0,A^0,H^\pm)$ not involved in $SU(2)_L$ breaking. The
lightest 
neutral scalar is 
basically the (real part of the) neutral component of the Higgs doublet 
involved in  $SU(2)_L$ breaking and has then a very small singlet component. 
The third neutral scalar has mass controlled by $M_{Z'}$ and is basically the 
singlet.

The lightest chargino is either
  predominantly a higgsino or a
gaugino.
In the neutralino sector, there is an extra $U(1)'$ zino and the 
higgsino ${\tilde S}$ as well as the four MSSM neutralinos. The mass matrix
 is 
$6\times 6$. For general values of the parameters  
the mass eigenstates will be 
complicated mixtures of higgsinos and gauginos.  
 Numerical examples of the pattern 
expected for charginos and neutralinos in different
 scenarios are given in~\cite{CDEEL}. 
 
The natural LSP candidate
is the lightest neutralino. In these models the LSP is usually
mostly $\tilde{B}$.
For large gaugino masses, however, the lightest 
neutralino is the singlino $\tilde{S}$, whose mass is 
of the order of $M_Z$.  It provides a  viable dark matter
candidate~\cite{decarlosespin}.

Masses for the squarks and sleptons 
can be  obtained 
directly from the MSSM formulae by setting $\mu\equiv h_ss/\sqrt{2}$
 and adding the 
pertinent $D$-term diagonal contributions from the $U(1)'$~\cite{dterms,lykken}. 
This extra term can produce significant mass deviations with 
respect to the minimal model and plays an important role
in the connection between parameters at the electroweak and string scales.



\subsection{Intermediate Scale Breaking}

A mechanism to generate intermediate scale $U(1)'$ breaking
in supersymmetric
theories utilizes the  $D$-flat directions~\cite{iscales}, 
 which are   in general present in the case of
  two or more SM  singlets with opposite signs of the $U(1)'$ charges.   
For simplicity, consider  {\it two}
 chiral multiplets $\hat{S}_{1,2}$  that are singlets
  under the SM  gauge group, with the $U(1)'$ charges
   $Q_{1S}$ and $Q_{2S}$, respectively.
If these charges have opposite signs ($Q_{1S}Q_{2S}<0$), there is a $D$-flat direction: 
\begin{equation}
Q_{1S} \langle S_1\rangle ^2+Q_{2S}\langle S_2 \rangle^2=0.
\end{equation}
In the absence of the self-coupling of $\hat{S}_1$ and $\hat{S}_2$ 
in the superpotential  there is an  $F$-flat  direction
in the ${ S}_{1,2}$ scalar field space as well.
 Then the only contribution to the scalar potential 
  along this $D$- and $F$-flat direction is 
due to the   soft mass-squared terms $m_{1S}^2|S_1|^2+m_{2S}^2|S_2|^2$. 
 For 
  the (real) component along the flat direction
   $s \equiv ({\sqrt{2|Q_{2S}|}}
 {\rm Re}    S_1 + {\sqrt{2|Q_{1S}|}}
 {\rm Re}    S_2 )/{\sqrt{|Q_{1S}|+|Q_{2S}|}}$
the potential is simply
\begin{equation}
V(s)=\frac{1}{2}m^2s^2,\ \  \ \  m^2\equiv
{{|Q_{2S}|m_{1S}^2+ |Q_{1S}|m_{2S}^2}\over{|Q_{1S}|+|Q_{2S}|}},
\label{msum}
\end{equation}
where $m_{1S,2S}^2$ are respectively the  $S_{1,2}$ soft mass-squared terms.  
$m^2$ is evaluated at the scale $s$.
 For 
$m^2$  positive at the 
string  scale it  can be driven to negative values at the 
electroweak scale if $\hat{S}_1$ and/or
$\hat{S}_2$ have a large
Yukawa coupling  to other fields,
 which is in general the case for this class of 
 superstring models, e.g.,  terms of the type
(\ref{sptrip}) with $h_D=\sqrt{2}g_U$ at $\MS$.  In this case, $V(s)$
 develops
a minimum along the flat direction and $s$ acquires a VEV.
   
 From the
minimization condition for (\ref{msum}) one sees that 
the VEV $\langle
s\rangle$ is determined by 
$\left. \left( m^2 +\frac{1}{2}
s\frac{dm^2}{d s}\right) \right|_{\langle s \rangle} = 0,$
which is satisfied very close to the scale 
$\mu_{ RAD}$ at 
which $m^2$ crosses
zero. This scale is fixed by the RGE evolution of
parameters from $\MS$ down to the electroweak scale and lies in general 
at an intermediate scale. It can be achieved with universal boundary
conditions at $\MS$ as long as there is a large Yukawa coupling of  SM singlets
to other matter, e.g., (\ref{sptrip}).  The precise value depends on the type
of couplings of $\hat{S}_{1,2}$  and the particle content of
 the model.  Examples  with  $\mu_{RAD}$ 
  in the range   $10^{15}$ GeV --  $10^{4}$ GeV 
 are  discussed in detail in~\cite{CCEEL} (see also~\cite{CL,SY}).

\subsubsection{Competition with Non-Renormalizable Operators}
 
The stabilization of the minimum along the $D$-flat direction
 can also be
due to non-renormalizable terms in the superpotential, which lift the
$F$-flat direction for sufficiently large  $s$. If these terms are 
important
below  $\mu_{ RAD}$, they will determine $\langle s\rangle$.
The relevant non-renormalizable terms
are of the form~\footnote{One can also 
have terms of the form
$\alpha_{K}^{K}\hat{S}^{2+K}\hat{\Phi}/{\MP^K}$,
where $\Phi$ is a SM singlet that does not acquire a VEV.
These have similar implications as the terms in (\ref{nrsup}).}
\begin{equation}
\label{nrsup}
W_{\rm NR}=\left(\frac{\alpha_{K}}{\MP}\right)^{K}\hat{S}^{3+K},
\end{equation}
where $\hat S$ is the (effective) chiral superfield  with the (real)
scalar component along the  $D$-flat
direction, $K=1,2,\cdots$,
and $\MP$ is the Planck scale. 

Including the $F$-term from (\ref{nrsup}), the potential along $s$ is
\begin{equation}
\label{potflat}
V(s)=\frac{1}{2}m^2s^2+{1\over 
{2(K+2)}}\left(\frac{s^{2+K}}{M^K}\right)^2,
\end{equation}
where $M \equiv C_K \MP/\alpha_{K}$, and $C_K$ is a coefficient of order unity.
The VEV of $s$ is
 then~\footnote{For simplicity,   soft-terms 
 of the type $(AW_{\rm NR}+{\mathrm H.c.})$ 
with $A\sim
m_{soft}$ are not included in (\ref{potflat}).
 Such terms do not affect the order
of magnitude estimates.} 
\begin{equation}
\label{vev}
\langle s\rangle=\left( |m| M^K \right)^\frac{1}{K+1}
\sim (m_{soft}M^K)^\frac{1}{K+1},
\end{equation}
where $m_{soft}={\cal O} (|m|)={\cal O} (M_Z)$ is a typical soft supersymmetry
breaking scale.
 $m^2$ is evaluated at the scale $\langle s\rangle$
and has to satisfy the necessary condition $m^2(\langle s\rangle)<0$. 
If non-renormalizable terms are negligible below $\mu_{ RAD}$, no solution
to (\ref{vev}) exists and $\langle s\rangle$ is fixed 
solely by the running $m^2$.

The coefficients $\alpha_{K}$ in (\ref{nrsup}) and thus $\MK$ in 
 (\ref{potflat})
are in principle calculable in 
a class of superstring models discussed above.
Depending on the $U(1)'$ charges and world-sheet 
symmetries of the superstring models, not all values of
$K$ are allowed. 
  It is expected that the  world-sheet integrals 
  that determine the  couplings of non-renormalizable terms
  are such that $M$ increases as $K$ increases.
 For  $K=1$ one  obtains (in a class of models):
 $M\sim 3\times 10^{17}$ GeV, 
and for $K=2$: $M\sim 7\times 10^{17}$ GeV.


\subsubsection{Higgs and Higgsino Mass Spectrum}

The mass of the physical field $s$ in the vacuum $\langle s\rangle$
is either   $\sim m_{soft}/4\pi$ for 
pure radiative breaking,
or  $\sim m_{soft}$ in 
 the case of stabilization by non-renormalizable terms.
Thus, the potential is very flat, with possible important cosmological
consequences.
The physical excitations along the transverse direction have an intermediate
mass scale.
There  remains one massless pseudoscalar,
 which  can acquire a mass from
 a soft supersymmetry breaking term of the
  type $AW_{\rm NR}$ or from  loop corrections. 

The fermionic part of the $Z'-S_1-S_2$ sector consists of three neutralinos
$(\tilde{B}',\tilde{S}_1,\tilde{S}_2)$.  One combination
is light, with mass of order $m_{soft}$ (if the minimum is fixed by
non-renormalizable terms) or of order $m_{soft}/4\pi$
 (obtained at one-loop order when 
 the minimum is instead determined by the
running of $m^2$).  The two other
neutralinos have  intermediate scale masses.

Other fields that  couple to $\hat{S}_{1,2}$ in the renormalizable
superpotential acquire intermediate scale masses, and those  that  do not
remain light.
The usual
 MSSM fields should belong to the
latter class. 

\subsubsection{$\mu$-Parameter}

The flat direction $S$ can have a set of non-renormalizable couplings
to MSSM states that offer a solution to the $\mu$ problem. 
The non-renormalizable $\mu$-generating terms are of the form,
\begin{equation}
\label{supermu}
W_{\mu}\sim \hat{H_1} \hat{H_2} \hat{S} \left( {\hat{S}\over \MK} \right)^P.
\label{nonrenmu}
\end{equation}
For breaking due to  non-renormalizable terms, $K=P$ yields  an effective
$\mu$-parameter
$\mu\sim
m_{soft}$, while for  pure radiative breaking  $\mu\sim
\mu_{RAD}^{P+1}/M^P$, which  depends crucially on the value of $\mu_{RAD}$.

\subsubsection{Fermion Masses}

Non-renormalizable couplings can also 
 yield mass hierarchies between  family generations. Generational up, down, and electron mass terms appear,
via
\begin{equation}
W_{u_i}\sim \hat{H}_2 \hat{Q}_i \hat{U}^c_i 
            \left( {\hat{S}\over \MK} \right)^{P^{'}_{u_i}}; 
W_{d_i}\sim \hat{H}_1 \hat{Q}_i \hat{D}^c_i 
            \left( {\hat{S}\over \MK} \right)^{P^{'}_{d_i}}; 
W_{e_i}\sim \hat{H}_1 \hat{L}_i \hat{E}^c_i 
            \left( {\hat{S}\over \MK} \right)^{P^{'}_{e_i}},
\label{nonrenql}
\end{equation}
with $i$ the family number.
$K$ and $P^{'}_i$ can be chosen to yield a realistic
 hierarchy for the first two generations~\cite{CCEEL}.
 Presumably the top mass is associated with a renormalizable coupling
 ($P'_{u_3}= 0$). The other third family masses
 do not fit  as well; it is  possible that $m_b$
 and $m_{\tau}$ are
 associated with some other mechanism, such as non-renormalizable
 operators involving
 the VEV of a different singlet.

Non-renormalizable terms yielding Majorana
 and Dirac neutrino terms may
also be present. They can   produce 
small (non-seesaw) Dirac masses. It is also possible to
have small Majorana
neutrino masses,  providing an interesting possibility for
  oscillations of the ordinary into sterile
neutrinos~\cite{PLprep}~\footnote{Another
mechanism for light sterile neutrinos from a $U(1)'$ is discussed
in~\cite{makeith}.}.
 A traditional seesaw can also be obtained, depending
 on the nature of the non-renormalizable 
operators (for details see~\cite{CCEEL}). 

\section{Conclusion}

The supersymmetric $Z'$ models  described above
(as motivated from a class of superstring models)
 provide a ``minimal'' extension of the 
MSSM. The electroweak breaking scenario
yields  phenomenologically acceptable 
gauge symmetry breaking patterns, which  involve a certain but not
excessive amount of 
fine-tuning. It predicts
interesting new phenomena  and can be tested at future colliders. 
The  intermediate scale scenario 
provides a framework in which 
intermediate scales can naturally occur,  with 
 interesting implications for
the $\mu$-parameter and the fermion  masses.

\subsubsection{Acknowledgments}

This work was supported by the U.S. Department of Energy Grant
 No. DOE-EY-76-02-3071 and NSF Career Advancement Award.
 We would like to thank F. del Aguila,
D.  Demir, G. Cleaver,
 J. R. Espinosa and L. Everett  for fruitful collaboration on 
 aspects of 
 $Z'$ physics  presented in this review.


\section*{References}

\begin{thebibliography}{99}
\bibitem{dterms} 
M. Drees, Phys. Lett. {\bf B181}, 279 (1986);
H. E. Haber and M. Sher, \PR{D35}{2206}{87};
J. R. Espinosa and M. Quir\'os, Phys. Lett. {\bf B279}, 92 (1992),
{\bf B302}, 51 (1993);
 G. Kane, C. Kolda, J. D. Wells, Phys. Rev.
Lett. {\bf 70}, 2686 (1993); 
D. Comelli and C. Verzegnassi, Phys. Lett. {\bf B303}, 277 (1993); 
Y. Kawamura and M. Tanaka, Prog. Th. Phys. {\bf 91}, 949 (1994).
H.-C. Cheng and L. Hall, Phys. Rev. {\bf D51}, 5289 (1995);
C. Kolda and S. Martin, Phys. Rev. {\bf D53}, 3871 (1996).
%
\bibitem{lykken} 
J. D. Lykken, hep-ph/9610218.
%
\bibitem{SY} D. Suematsu and Y. Yamagishi, Int. J. Mod. Phys. {\bf  
A10}, 4521 (1995).
%
\bibitem{CL} M. Cveti\v c and P. Langacker, Phys. Rev. {\bf D54},  
3570 (1996) and Mod. Phys. Lett. {\bf 11A}, 1247 (1996).
%
\bibitem{CDEEL} M. Cveti\v c, D. A. Demir,
J. R. Espinosa, L. Everett and P.
Langacker, hep-ph/9703317.
%
\bibitem{CCEEL} 
G. Cleaver, M. Cveti\v c, J. R. Espinosa, L. Everett and
P. Langacker, hep-ph/9705391.
%
\bibitem{GRH}
J. F. Gunion, L. Roszkowski, and H. E. Haber, \PL{B189}{409}{87};
   \PR{D38}{105}{88}.
%
\bibitem{keithma}
E. Keith, E. Ma, and B. Mukhopadhyaya, \PR{D55}{3111}{97};
E. Keith and E. Ma, hep-ph/9704441.
%
\bibitem{jhtr}
For a review, see J. Hewett and T. Rizzo, \Rep{183}{193}{89}.
%
\bibitem{iscales} M. Dine, V. Kaplunovsky, M. Mangano, C. Nappi and N. Seiberg, 
Nucl. Phys. {\bf B259}, 549 (1985);
M. Mangano, \ZP{C28}{613}{85};
G.~Costa, F.~Feruglio, F.~Gabbiani and F.~Zwirner,
Nucl. Phys. {\bf B286}, 325 (1987);
J. Ellis, K. Enqvist, D. V. Nanopoulos and K. Olive, Phys. Lett. {\bf 188B}, 415 
(1987);
R. Arnowitt and P. Nath, Phys. Rev. Lett. {\bf 60}, 1817 (1988).
%
%
\bibitem{ABK}{I. Antoniadis, C. Bachas and C. Kounnas, Nucl. Phys. {\bf B289},
 87 (1987); H. Kawai, D. Lewellen and S. H. H. Tye, Phys. Rev. Lett. {\bf 57},
 1832 (1986) and  Phys. Rev. {\bf D34}, 3794 (1986).}
%
%
\bibitem{NAHE}{I. Antoniadis, J. Ellis, J. Hagelin and D. V. Nanopoulos, Phys.
Lett. {\bf B231}, 65 (1989).}
%
\bibitem{faraggi90a}{A. Faraggi, D. V. Nanopoulos and K. Yuan, 
Nucl. Phys. {\bf B335}, 347 (1990); 
A. Faraggi, Phys. Lett. {\bf B278}, 131 (1992).}
%
\bibitem{CHL}{S. Chaudhuri, S.-W. Chung, G. Hockney and J. Lykken,
Nucl. Phys. {\bf B456}, 89 (1995); S. Chaudhuri,  G. Hockney and J. Lykken,
Nucl. Phys. {\bf B469}, 357 (1996).}
%

\bibitem{dienes} For a review, see K. R. Dienes, hep-th/9602045.
%
\bibitem{INQ}{ L. E. Ib\'a\~nez, J. E. Kim, H. P. Nilles, and F. Quevedo,
 Phys. Lett. {\bf 191B}, 282 (1987); 
 A. Font, L. E. Ib\'a\~nez, H. P. Nilles and F. Quevedo,
 Phys. Lett. {\bf 210B}, 101 (1988), {\it  erratum, ibid.},
 {\bf B213}, 564 (1988).} 
%
\bibitem{AFIU}
G. Aldazabal, A. Font, L. E. Ib\'a\~nez, and A.M. Uranga, \NP{B452}{3}{95} and  
\NP{B465}{34}{96}.


\bibitem{Kakushadze}{
Z. Kakushadze and  S. H. H.  Tye, Phys. Lett. {\bf B392}, 335 (1997),
 Phys. Rev. {\bf D55}, 7878 (1997)  and
  Phys.  Rev. {\bf D55}, 7896 (1997).}

%
\bibitem{MUPROB}{J. E. Kim and H. P. Nilles, Phys. Lett.
{\bf B138}, 150 (1984).}
%
\bibitem{esp}
M. Pietroni, \NP{B402}{27}{93};
J. R. Espinosa, in preparation.
%
\bibitem{COSTRINGS} 
R. Brandenberger, A.-C. Davis, and
M. Rees, Phys. Lett. {\bf B349}, 329 (1995), and references therein.
%
\bibitem{leptonasym} 
           {M. Fukugita and T. Yanagida, Phys. Lett. {\bf B174},
           45 (1986);  Phys. Rev. {\bf D42}, 1285 (1990);
           P. Langacker, R.D. Peccei and T. Yanagida, Mod.
           Phys. Lett. {\bf A1}, 541 (1986); W. Buchm\"{u}ller and
           M. Pl\"{u}macher, \PL{B389}{73}{96}.}
\bibitem{buch}{
           W. Buchm\"{u}ller and T. Yanagida, Phys. Lett.
           {\bf B302}, 240 (1993).}
%
\bibitem{decarlosespin} 
  B. de Carlos and J. R. Espinosa,
   hep-ph/9705315.
%
\bibitem{CG}{For a review see, {\it e.g.,} 
M. Cveti\v c and S. Godfrey, in Proceedings of  {\it Electroweak Symmetry 
Breaking
and Beyond the Standard Model}, eds. T. Barklow, S Dawson, H. Haber and J.
Siegrist (World Scientific 1995), hep-ph/9504216, and references therein.}
%
\bibitem{jepl}{See
P. Langacker, p 883 of~\cite{precision}.
The numbers presented here are based on
J. Erler and P. Langacker, {\it Proceedings of the Ringberg
Workshop on the Higgs Puzzle}, hep-ph/9703428.}
%
\bibitem{precision}{
{\it Precision Tests of the Standard
Electroweak Model}, ed. P. Langacker (World, Singapore, 1995).}
%
\bibitem{dl}
L.S. Durkin and P. Langacker, \PL{B166}{436}{86}.
%
\bibitem{robinett}
R.~Robinett, Phys. Rev. {\bf D26}, 2388 (1982);
  R.~Robinett and J.~Rosner, Phys. Rev. {\bf D25}, 3036 (1982) and
 {\bf D26}, 2396 (1982).
%
\bibitem{etamodel}
E. Witten, \NP{B258}{75}{85}.
%
\bibitem{lrmodels}
For a review see R.N. Mohapatra, {\sl
Unification and Supersymmetry} (Springer, New York, 1986);
for constraints on general LR  models, see
P. Langacker and S. Uma Sankar, \PL{D40}{1569}{89}.

%
\bibitem{amaldi}
U. Amaldi \etal, \PR{D36}{1385}{87}.
%
\bibitem{kayser}
M.~Cveti\v c, B.~Kayser, and P.~Langacker,
Phys. Rev. Lett. {\bf 68}, 2871 (1992).
%
%
\bibitem{heavylep} 
P. Chiappetta \etal, Phys. Rev.
{\bf D54}, 789 (1996);
G. Altarelli \etal, Phys. Lett. {\bf B375}, 292 (1996).
%
\bibitem{lightlep} 
V. Barger, K. Cheung, and P. Langacker, 
Phys. Lett. {\bf B381}, 226 (1996).
%
\bibitem{leptoconstraints} 
K. Agashe \etal, Phys. Lett. {\bf B385}, 218 (1996).
%
\bibitem{lepdata}{The current situation is summarized in the
joint report of the LEP Collaborations, LEP Electroweak
Working Group, and SLD Heavy Flavor Group, CERN-PPE/96-183.}
%
\bibitem{bkm} 
K. S. Babu, C. Kolda, and  J. March-Russell,
Phys. Rev. {\bf D54}, 4635 (1996).
%
\bibitem{stringylepto} 
A. E. Faraggi and M. Masip, \PL{B388}{524}{96};
J. L. Lopez and D. V. Nanopoulos, \PR{D55}{397}{97}.
%
\bibitem{thirdfam} 
B. Holdom, \PL{B339}{114}{94}, \con{B351}{279}{95};
P. Frampton, M. Wise, and B. Wright, 
Phys. Rev. {\bf D54}, 5820 (1996).
%
\bibitem{fermio} 
A. Donini \etal, hep-ph/9705450.
%
\bibitem{kmix1}
B. Holdom, \PL{B166}{196}{86}.
%
\bibitem{kmix2}
F. del Aguila \etal, \NP{B283}{50}{87};
F. del Aguila, G. D. Coughlan and M. Quir\'os; Nucl. Phys. {\bf
B307}, 633 (1988), [E-{\bf B312}, 751 (1989)]; F.~del~Aguila, M.~Masip and
M.~P\'erez-Victoria, Nucl. Phys. {\bf B456}, 531 (1995).
%
\bibitem{ACLIV}
F. del Aguila, M. Cveti\v c and P. Langacker,
\PR{D52}{37}{95}.
%
\bibitem{mixdan} K. R.~Dienes, C.~Kolda and J.~March-Russell,
\NP{B492}{104}{97}.
%
\bibitem{leptolimits}
See the Appendix of the second paper in~\cite{CL}.
%
\bibitem{nonunivz}
Y.~Zhang and B. L.~Young, Phys. Rev. {\bf D51}, 6584 (1995);
B. Holdom and M. V. Ramana, Phys. Lett. {\bf B365}, 309 (1996).
%
\bibitem{london} 
P. Langacker and D. London, Phys. Rev. {\bf D38}, 886 (1988), and
D. London, p 951 of~\cite{precision}.
%
\bibitem{nardi} 
E. Nardi, \PR{D48}{1240}{93}.
%
\bibitem{suematsu} 
D. Suematsu, hep-ph/9705405.
%
\bibitem{colliderlim} 
CDF: F. Abe \etal, FERMILAB-PUB-97-122-E;
D0: S. Abachi \etal, \PL{B385}{471}{96}.
%
\bibitem{exoticdecay} 
V. Barger \etal, \PR{D35}{2893}{87}.
%
\bibitem{gkk} 
T. Gherghetta, T. A. Kaeding, and G. L. Kane, hep-ph/9701343.
%
\bibitem{jlr} 
J. L. Rosner, \PL{B387}{113}{96}.
%
\bibitem{gg} 
H. Georgi and S. L. Glashow, \PL{B387}{341}{96}.
%
\bibitem{GHP} 
S. Godfrey, J. L. Hewett, and L. E. Price, hep-ph/9704291.
%
\bibitem{sg} 
S. Godfrey, \PR{D51}{1402}{95}.
%
\bibitem{e+e-}
A. Djouadi, A. Leike, T. Riemann, D. Schaile and C. Verzegnassi,
Z. Phys. {\bf C56} 289 (1992);
A. Leike, Z. Phys. {\bf C62}, 265 (1994);
D. Choudhury, F. Cuypers, and A. Leike, Phys. Lett. {\bf B333}
531 (1994).
For earlier references, see~\cite{CG}.
%
\bibitem{LRR}{P. Langacker, R. W. Robinett, and J. L. Rosner, Phys.  
Rev. {\bf D30}, 1470 (1984).}
%
\bibitem{dittmar}
M. Dittmar, \PR{D55}{161}{97}.
%
\bibitem{ACL}
{F. del Aguila, M.
Cveti\v c, and P. Langacker, Phys. Rev. {\bf D48}, R969 (1993).}
%
\bibitem{FT}
{ A. Fiandrino and  P. Taxil, Phys. Rev. {\bf D44}, 3490 (1991);
 Phys. Lett. {\bf B292}, 242 (1992).}
%
\bibitem{AAC}
{J. Anderson, M. Austern, and R. N. Cahn,  Phys.
Rev. Lett. {\bf 69}, 25 (1992) and Phys. Rev. {\bf D46}, 290 (1992).}
%
\bibitem{RM}
T. G. Rizzo, Phys. Rev. {\bf D48}, 4236 (1993).
%
\bibitem{MOH}
{P. Mohapatra,  Mod. Phys. Lett. {\bf A8}, 771 (1993).}
%
\bibitem{RI}
T. G. Rizzo, Phys. Lett.  {\bf B192}, 125 (1987);
J. L. Hewett and T. G. Rizzo, Phys. Rev. {\bf D45}, 161 (1992).
%
\bibitem{CLII}
{M.~Cveti\v c and P. Langacker,  Phys. Rev. {\bf D46}, R14 (1992).}
%
\bibitem{CLIV}
{M.~Cveti\v c and P. Langacker,  Phys. Rev. {\bf D46}, 4943 (1992).}
%
\bibitem{RII}
{T. G. Rizzo, Phys. Rev. {\bf D47}, 956 (1993).}
%
\bibitem{ACLIII}
{F. del Aguila and  M. Cveti\v c,  Phys. Rev. {\bf D50}, 3158 (1994).}
%
\bibitem{HREE}
{J. Hewett and T. Rizzo, in the  Proceedings of the {\it
Workshop on Physics and
Experiments with Linear $e^+e^-$ Colliders},  September 1991,  
Saariselke\" a,
Finland, R. Orava ed., Vol. II, p. 489; {\it ibidem} p. 501.}
%
%
\bibitem{exotic}{For indirect constraints, see~\cite{london}.
For a recent discussion,
including direct collider constraints, see V. Barger, M. S. Berger,
and R. J. N. Phillips, Phys. Rev. {\bf D52}, 1663 (1995).}
%
\bibitem{nrl}{E. Nardi, E. Roulet, and D. Tommasini,
Nucl Phys. {\bf B386}, 239 (1992) and Phys. Rev. {\bf D46}, 3040 (1992).}
%
%
\bibitem{Langacker}{P. Langacker, Phys. Rep. {\bf 72}, 185 (1981).}
%
\bibitem{SUSYGUTS}{S. Dimopoulos, S. Raby and F. Wilczek, Phys. Rev. {\bf D24},
1681 (1981);
L. E. Ib\'a\~ nez and G. G. Ross,
Phys. Lett. {\bf 105B}, 439 (1981).  }
%
\bibitem{LP}{For recent discussions, see
P. Langacker and N. Polonsky, Phys. Rev. {\bf D52}, 3081 (1985);
P. Langacker, in Proceedings of {\it SUSY-95}, hep-ph/9511207, 
and references
therein.}
%
\bibitem{ALS}{
For  a review, see
the article by
I. Hinchliffe on {\it Quantum Chromodynamics}, in~\cite{barnett96};
M. Shifman,  Mod. Phys. Lett. {\bf A10}, 605 (1995);
J. Erler and P. Langacker, Phys. Rev. {\bf D52}, 441 (1995) and~\cite{jepl}.}
%

\bibitem{barnett96}{R. M.
Barnett {\it \etal, Reviews of Particle Physics}, Phys.
Rev. {\bf D54}, 1 (1996).}
%
\bibitem{DW}{S. Dimopoulos and F. Wilczek, Santa Barbara preprint
 81-0600 (1981).}
%
\bibitem{Nilles}{For a review, see, e.g., H. P. Nilles,
Phys. Rep. {\bf 110}, 1 (1984), and references therein.}
%

\bibitem{AM}{C.
S. Aulakh and R. N. Mohapatra, Phys. Rev. {\bf D28}, 217 (1983).}
\bibitem{K}{V. Kaplunovsky, Nucl. Phys. {\bf B307}, 145  (1988), {\it erratum},
{\it ibid.} {\bf B383}, 436 (1992).}
%
\bibitem{BL}{See e.g.,
K. R. Dienes and A. E. Faraggi,  Phys. Rev. Lett. {\bf 75}, 2646 (1995);
K. R.  Dienes, Ref.~\cite{dienes};
 P. Binetruy and P. Langacker, in preparation.} 
%

\bibitem{WittenII}{E. Witten, Nucl. Phys. {\bf B471}, 135 (1996).} 

%
\bibitem{S}{A. N. Schellekens, Phys. Lett. {\bf B237}, 363 (1990).}
%
\bibitem{dudas}
For recent work, see e.g., P. Binetruy and E. Dudas,
\PL{B389}{503}{96} and references therein.
%
\bibitem{DSW}{M. Dine, N. Seiberg and E. Witten, Nucl. Phys. {\bf B289}, 
585 (1986); J. Atick,
L. Dixon and A. Sen, Nucl. Phys. {\bf B292}, 109 (1987). }

%
\bibitem{DKI}{L. Dixon and V. Kaplunovsky, unpublished.} 
%
\bibitem{shrock}
{V. Jain and R. Shrock, Phys. Lett. {\bf B352}, 83 (1995)
and hep-ph/9507238;
 Y. Nir, Phys. Lett. {\bf B354}, 107
(1995).}
%
\bibitem{F}{A. E. Faraggi,  Phys. Lett. {\bf
B274}, 47 (1992) and Nucl. Phys. {\bf B403}, 102 (1993)}
%
\bibitem{binetruy96}{
P. Binetruy, N. Irges, S. Lavignac and P. Ramond, \PL{B403}{38}{97};
P. Binetruy, S. Lavignac and P. Ramond, \NP{B477}{353}{96}; 
J. K. Elwood, N. Irges and P. Ramond, hep-ph/9705270.}
%

\bibitem{KLS} {A. Kusenko, P. Langacker and G. Segr\`e, Phys. Rev. {\bf D54}, 
 5824 (1996); A. Kusenko and P. Langacker, Phys. Lett. {\bf B391},  29 (1997).
 }
%
\bibitem{Luo}{We thank M.-X. Luo for pointing that out to us.}
%
\bibitem{PLprep}{P. Langacker, in preparation.}
%
\bibitem{makeith}{E. Ma, Mod. Phys. Lett. {\bf A11}, 1893 (1996); E. Keith and
E. Ma, hep-ph/9603353.}
%
\end{thebibliography}
\end{document}